# Wave propagation in highly inhomogeneous thin films: exactly solvable models

Guillaume Petite[(1)] and Alexander B. Shvartsburg[(2)]
(1) Laboratoire des Solides Irradiés, UMR 7642, CEA-DSM, CNRS et Ecole Polytechnique,
F-91128, Palaiseau, France
(2) Central Design Bureau for Unique Instrumentation of the Russian Academy of Sciences
Butlerov Str. 15,  Moscow, Russian Federation

**Abstract :** We present an exact treatment of wave propagation in some inhomogeneous thin films with highly space-dependent dielectric constant. It is based on a space transformation which replaces the physical space by the optical path. In the new space, the dispersion equation is that of a normal progressive wave. We will show that the dispersion properties of such films are plasma- or waveguide-like, the characteristic frequency being determined by the spatial characteristics of the dielectric constant's variations only. The theory is scalable, so that it can be applied in any wavelength range : optical, IR, radiofrequency, etc. depending only on the characteristic space scales.  Several applications will be presented, concerning the reflection properties of such films (broadband anti-reflection, or dichroic coatings) or to the propagation and transmission through the film. We will show that depending on the type of space dependence, an incident wave can either propagate or tunnel through such films. We will investigate the behaviour of the light group-velocity and tunneling time inside or through such films. Though we can reproduce the phase-time saturation corresponding to the Hartman effect, analysis of the group velocity in the tunneling case shows no sign of superluminal propagation. A strong frequency dependence can be obtained in some situations, which allows to anticipate a strong reshaping of brodband laser pulses.

## 1. INTRODUCTION

The problem of wave propagation in media with rapidly varying dielectric constants has received attention since the early stage of electromagnetism. The first analytical results were obtained by Rayleigh for waves whose velocity inside the medium depends linearly upon the coordinate [(1)]. Later the linear profile for $\varepsilon(z)$[(2)] - z being the direction of propagation of the light - as well as an exponential and more general Epstein profiles[(3)] were used for the analysis of radio propagation in the ionosphere. Some more complicated distributions were modeled by piece-wise profiles of $\varepsilon(z)$[(4)], described by a WKB approximation[(5)] or treated numerically[(6)]. These researches focused on the propagation of EM waves in heterogeneous media with positive $\varepsilon$, although the tunneling phenomena, which arise when $\varepsilon < 0$, were touched sometimes, e.g., in the case of radio waves percolation nearby the ionospheric maxima[(7)].

Modern technologies now allow to realize thin films with a significant dielectric constant variation over a length of the order or even smaller than visible wavelengths. Such thin films of, e.g., ZnSe or silicon oxynitride, were shown to provide broadband antireflection properties (Sankur *et al.*[(8)]). Methods for real-time monitoring and control of the growth of transparent inhomogeneous layers, based on reflectometry and ellipsometry of growing layers, were elaborated (Kildemo *et al* [(9)]). However, the profiles realized in this way are much more general than the ones that received analytical solutions, and numerical treatments[(6)] generally fail revealing the essence of the physics of wave propagation in such objects. In this paper, we summarize the work we performed on this topic [(10,11)] which is based on the following strategy :
- using profiles of $\varepsilon(z)$ containing enough free parameters to encompass most of the general features of the films that can be generated (e.g. concave or convex variations of $\varepsilon(z)$, monotonous or not)
- the material of the film, as well as of the eventual substrate has no intrinsic dispersion, so that any observed dispersion will find its origin in the spatial variations of the dielectric constant, hereafter referred to as "Heterogeneity Induced Dispersion" (HID)

- relying only on analytical solutions of the wave equation (which will be obtained in a transformed space) in order to keep a good insight on the origin of the observed phenomena
- though we will explicitly refer to the optical domain, the theory can be spatially scaled so that the conclusions apply to any field of electromagnetism

The properties that we are interested in are the reflection properties of the films, as well as its transmission properties. In particular, we will consider the problems of tunneling through films which cannot support field propagation. Indeed the advent of lasers attracted attention upon light tunneling in a series of optoelectronics problems, such as, e.g., the evanescent modes in dielectric waveguides[12], surface waves on microspheres[13], Goos – Hanchen effect for optical coatings[14]. A new burst of interest into these phenomena was stimulated by the intriguing perspective of superluminal light propagation through opaque barriers[15]. The experiments in microwave range with "undersized" waveguides[16] and bi-prism devices[17] as well as the analysis of spatial displacement of the peak of a tunneling pulse[18] and the direct measurement of photons tunneling time[19,20] were considered by some authors in favor of the concept of superluminal phase time for the tunneling electromagnetic waves[21]. However, this concept aroused controversial viewpoints[22,23].

Finally, we will see that some of the studied films can present very strong dispersion properties. Obviously, and though we do not explicitly treat this problem, this will induce a very strong reshaping of broadband ultrashort pulses propagating through such films, and could eventually offer the opportunity of manipulating their shape in a controlled way.

## 2. HETEROGENEITY INDUCED DISPERSION (HID) OF THIN DIELECTRIC FILMS.

Let us consider an inhomogeneous dielectric film as a plane dielectric layer with thickness $d$ and dielectric susceptibility $\varepsilon(z)$, $0 \leq z \leq d$. A linearly polarized EM wave, that we assume to propagate in the z-direction (normal incidence conditions), is described by Maxwell equations, linking the $E_x$ and $H_y$ components of the wave :

$$\frac{\partial E_x}{\partial z} = -\frac{1}{c}\frac{\partial H_y}{\partial t} \tag{1}$$

$$\frac{\partial H_y}{\partial z} = -\frac{\varepsilon(z)}{c}\frac{\partial E_x}{\partial t} \tag{2}$$

We will use the following model of dielectric susceptibility profile $\varepsilon(z)$

$$\varepsilon(z) = n_0^2 U^2(z) \quad , \quad \text{with} \quad U(z) = \left(1 + \frac{s_1 z}{L_1} + \frac{s_2 z^2}{L_2^2}\right)^{-1} \; ; \; s_1 = 0, \pm 1 \; ; \; s_2 = 0, \pm 1 \tag{3}$$

Here $n_0$ is the refractive index value on the interface z = 0; the distribution (3) is considered in the region z ≥ 0. The characteristic spatial scales $L_1$ and $L_2$ as well as the values $s_1$ and $s_2$ are the free parameters of model (3). Note that the Rayleigh profile of eq. (1) above corresponds to the limit of $U(t)$ when the scale $L_2 \to \infty$. The EM field in normal incidence can be described with help of a single component vector-potential cast under the form $\mathbf{A}(z,t) = \mathbf{A_0} \psi(z,t)$, where $\psi$ is a scalar function such that (for convenience, we put $A_0=1$)

$$E_x = -\frac{1}{c}\frac{\partial \psi}{\partial t} \; ; \; H_y = \frac{\partial \psi}{\partial z} \tag{4}$$

which allows to reduce the system (1)-(2) to the single equation

$$\frac{\partial^2 \psi}{\partial z^2} - \frac{n_0^2 U^2(z)}{c^2}\frac{\partial^2 \psi}{\partial t^2} = 0 \tag{5}$$

which does not admit obvious solutions. Using a new function $F$ and a new variable $\eta$

$$F = \psi\sqrt{U(z)} \quad ; \quad \eta = \int_0^z U(\zeta)d\zeta \tag{6}$$

and using (3) transforms eq.(5) into a new equation with constant coefficients

$$\frac{\partial^2 F}{\partial \eta^2} - \frac{n_0^2}{c^2}\frac{\partial^2 F}{\partial t^2} = p^2 F \tag{7}$$

$$p^2 = \frac{1}{4L_1^2} - \frac{s_2}{L_2^2} \tag{8}$$

A solution of eq.(7) can be built by superposition of waves with wavenumber q, traveling in the $\eta$ - direction

$$F = \exp i(q\eta - \omega t) \quad \text{with} \quad q = \frac{\omega n_0}{c} N \quad ; \quad N = \sqrt{1 - \frac{c^2 p^2}{n_0^2 \omega^2}} \tag{9}$$

At this point, we note that solutions in the form of traveling waves are obtained only if the expression under the radical is positive. This is always the case if $p^2<0$, i.e. for $s_2=+1$, if $L_2>2L_1$. In the opposite case ($p^2>0$), the availability of traveling wave solutions is subject to a condition concerning $\omega$, which writes

$$\omega > \Omega \quad \text{with} \quad \Omega = \frac{c\sqrt{y^2 - s_2}}{n_0 L_2} \quad \text{and} \quad y = L_2/2L_1 \tag{10}$$

Combining (9) and (6), we obtain the function $\psi$ determining the vector-potential; whose substitution into (4) brings the explicit expressions for the field components

$$E_x = \frac{i\omega}{c\sqrt{U}}\exp i(q\eta - \omega\tau) \tag{11}$$

$$H_y = \frac{i\omega n_0 \sqrt{U}}{c}(N - iG)\exp i(q\eta - \omega\tau) \tag{12}$$

with

$$G = \frac{cs_1}{2\omega n_0 L_1}\left(1 + \frac{2s_2 z L_1}{s_1 L_2^2}\right) \tag{13}$$

Thus we found an exact solution describing the EM wave in an inhomogeneous layer (3). This solution can be used for finding the reflection coefficient of the wave, incidenting normally from the vacuum on the layer's interface $z = 0$, and to study the propagation of tunneling of waves though such layers.

## 3. ANTIREFLEXION PROPERTIES INDUCED BY HID

The reflexion properties of such films were investigated in detail in [10]. We present here the essential results of this study. In all cases we consider here a wavelength range in which propagation in the film is possible. In the case of films presenting a cut-off frequency condition (10) must be realized. We first illustrate the "intrinsic" properties of such films – considered as "self-standing" in vacuum - of different types and show their dependence on the type of profile.

Figure 1 illustrates the case of a film with $p^2>0$, $s_1>0$, $s_2<0$, $y=1$, $n_0 = 1.73$ and of different thickness characterized by the parameter $\alpha=d/L_2$, where $d$ is the film thickness. Under such conditions, with $\alpha$ ranging from 0.1 to 0.3, the dielectric constant is continuously decreasing throughout the film. Such a film possess a cut-off frequency given by :

$$\Omega_1 = \frac{c\sqrt{1+y^2}}{n_0 L_2} \tag{14}$$

We see that such a film has a low intensity reflection coefficient over a very large frequency range (typically less than 5% in the frequency range $\Omega$ to $2\Omega$). Note that the frequency is expressed in reduced coordinates : due to the definition of $\Omega$ (10), these antireflection properties can be obtained in any frequency range, varying only the length scales $L_2$ and $L_1$ (which are linked by the value of $y$). In the case $L_2 = 200$ nm, the cut-off frequency $\Omega$ is 1.22 $10^{15}$ rad.s$^{-1}$: according to the curve, related to the value $\alpha = 0.15$, one can find, e.g., that the reflectivity of a film with thickness $d = 30$ nm in a spectral range 0.5 µm < $\lambda$ < 1.55 µm does not exceed 5%. The same curve shows that a film with the same characteristics, but with $L_2 = 2$ µm and 10 times thicker, will exhibit the same antireflection properties in the far IR range (5 µm < $\lambda$ < 15.5 µm).

Figure 2 illustrates the case of films with $p^2<0$, which do not have a cut-off frequency. Here $s_2$ is positive, and $y=0.75$. In this case, the dielectric constant is continuously increasing throughout the film. Here the "characteristic frequency" used to define the reduced frequency $x_2$ writes

$$\Omega_2 = \frac{c\sqrt{1-y^2}}{n_0 L_2} \tag{15}$$

thus differing somewhat from definition 10 in order to be made real, but it does not have as above the physical meaning of a cut-off frequency. We see here that such films present a quite different behavior, clearly visible in the case of the film with $\alpha=1$ : this film presents a maximum of reflectivity for a reduced frequency of 1.5, and a minimum of reflection for a reduced frequency of 3, i.e. they present a rather strong dichroic character.

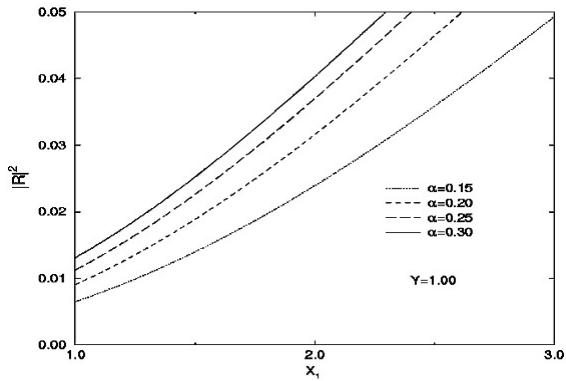

**Figure 1 :** Antireflection properties of thin inhomogeneous dielectric film in the case $s_2 <0$ and $p^2 > 0$ (8). The reflection coefficient |R|$^2$ is plotted vs the normalized frequency $x_1=\omega/\Omega_1$, for $n_0 = 1.73$, and for different values of the parameter $\alpha =$

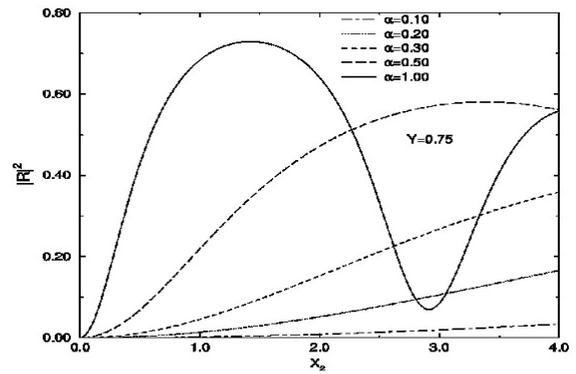

**Figure 2** : Spectra of reflectivity of inhomogeneous film in a case $s_2>0$ and $p^2 < 0$ (8); The reflection coefficient |R|$^2$ is plotted vs. the normalized frequency $x_2=\omega/\Omega_2$ for $n_0 = 1.73$, and for different values of the parameter $\alpha = d/L_2$.

$d/L_2$.

Now such films are seldom used as self-standing (at least in the optical domain, but it could be so in the RF one) and are usually applied as coatings. Figure 3 considers such a case, where an antireflection coating analogous to that of fig. 1, but with *y*=0.25 has been applied on a dielectric substrate with a complex dielectric constant $\varepsilon_2=(n_2+\chi_2)^2$, with $n_2$=3.5 and $\chi_2$=0.75. Such a lossy dielectric presents in itself a rather strong reflection coefficient (typically 0.32). The thin film acts as an antireflexion coating in a broad frequency region which can be chosen by adjusting the two length scales of the profile $U(z)$ and its thickness, i.e. the parameter $\alpha$.

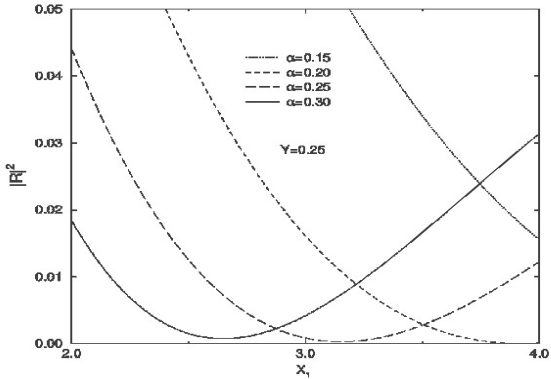

**Figure 3** : Broadband antireflection properties of a thin film, coating a lossy dielectric with $n_2$ = 3.5, $\chi_2$ = 0.7; the dependence of the reflectivity |R|² upon the normalized frequency $x_1 = \omega/\Omega_1$, relating to the case $y$ = 0.25, $n_0$ = 1.6, $p^2$>0, $s_2$ <0, is represented for different values of the parameter $\alpha = d/L_2$. The same dielectric without coating has a reflection coefficient |R|² = 0 .32

In all the above examples, the gradient of dielectric susceptibility strongly influences the frequency dispersion of the film's reflectivity. Some key properties of this effect should be pointed out, namely :
- A controlled formation of the areas of both positive and negative inhomogeneity-induced dispersion of a dielectric film which can be provided in an arbitrary spectral range via an appropriate choice of the dielectric susceptibility profile $U(z)$.
- Unlike the natural dispersion of materials, where a strong dispersion nearby a resonance is accompanied by an enhanced absorption, the artificial inhomogeneity-induced dispersion can be formed in a spectral range which is far from the absorption bands of the material.
- The thickness of an inhomogeneous antireflection film may be several times smaller than that of a standard homogeneous quarter-wave plate for a given wavelength, and even more than that of more complex multilayer coatings commonly used. For instance, considering a wavelength of 10.2 µm passing through the film relating to figure 3, we obtain a thickness of 0.57 µm, instead of 1.5 µm for a standard plate with the same value of $n_0$ =1.6. This difference may be useful for optimizing the sizes of antireflection systems.

## 4. PROPAGATION OR TUNNELING OF WAVES THROUGH CONVEX OR CONCAVE BARRIERS

We now want to analyze in more detail the question of the propagation (or in some case tunneling) of waves through films with profiles of their dielectric constants defined in (3). Hereafter we will consider the case of symmetric profiles, with opposite signs for $s_1$ and $s_2$. In such a case, quantities $L_1$, $L_2$ and $d$ are linked by the relation $L_1 = d/4y^2$, $L_2 = d/2y$. Two such classes of films can be defined presenting either a maximum ($s_1$ = -1, $s_2$ = +1, convex profile) or a minimum ($s_1$ =+1, $s_2$ = -1, concave profile) with a value $U_m$

$$U_m = \left(1 + s_1 y^2\right)^{-1} \qquad (16)$$

The characteristic frequencies (10) are different for concave ($\Omega_1$) and convex ($\Omega_2$) profiles :

$$\Omega_1^2 = \frac{c^2(1+y^2)}{n_0^2 L_2^2}; \qquad \Omega_2^2 = \frac{c^2(y^2-1)}{n_0^2 L_2^2} \qquad (17)$$

Only the concave profiles possess a cut-off frequency below which the wave can only tunnel through the film, a situation that will be studied later.

**4.1. - Group velocities of traveling EM waves in concave and convex photonic barriers.**

The spatial waveforms of the EM field inside the heterogeneous medium are non – sinusoidal and this field is formed due to interference of forward and backward waves. Hence the group velocities $v_g$ of these waveforms have to be found by means of energy flux (Poynting vector) **P** and energy density $W$[7]:

$$\mathbf{v}_g = \frac{\mathbf{P}}{W}; \quad \mathbf{P} = \frac{c}{4\pi}\mathrm{Re}\left[\mathbf{E}\wedge\mathbf{H}^*\right] \; ; \quad W = \frac{1}{8\pi}\left(\varepsilon|\mathbf{E}|^2 + |\mathbf{H}|^2\right) \qquad (18)$$

The spatial structure of the EM field inside the barrier is formed by the interference of forward wave, passing through the plane $z = 0$, and backward one, reflected from the plane $z = d$. Using formulae (11) – (12) one can present these waves in a form

$$E_x = \frac{i\omega C}{c}\left(e^{iq\eta} + Qe^{-iq\eta}\right)\frac{1}{\sqrt{U}}$$

$$H_y = iqC\sqrt{U}\left[\frac{iU_z}{2qU^2}\left(e^{iq\eta} + Qe^{-iq\eta}\right) + e^{iq\eta} - Qe^{-iq\eta}\right] \qquad (19)$$

with $U_z = -\frac{2U^2}{qL_2}\left(ys_1 + \frac{zs_2}{L_2}\right)$ (20)

For simplicity the time – dependent factor $\exp(-i\omega t)$ is omitted here and below. The dimensionless parameter $Q$ describes the reflectivity of the far boundary $z = d$. Introducing the reflection coefficient of the film $R$, we can write the continuity conditions on the plane $z = 0$ ($E_i$ is the electric component of the incidenting wave):

$$E_i(1+R) = \frac{i\omega C}{c}(1+Q) \qquad (21)$$

$$E_i(1-R) = \frac{i\omega n_0 NC}{c}\left[-\frac{is_1}{2qL_1}(1+Q) + 1 - Q\right] \qquad (22)$$

$$C = \frac{-icE_i(1+R)}{1+Q} \qquad (23)$$

and parameter $Q$ can be found from the continuity conditions on the interface $z = d$:

$$Q = \frac{-\exp(2iq\eta_0)\left(1 - \frac{i}{2}s_1\gamma - n_0 N\right)}{1 - \frac{i}{2}s_1\gamma + n_0 N} \qquad \gamma = c/\omega L_1 \qquad (24)$$

Expressions from which the fields, and then the Poynting vector and the energy density can be found through some long but easy algebra[11] allowing to express in each point $z$ the group velocity of the wave as

$$v_g(z) = \frac{4c}{U(z)\theta_+(z)} \qquad (25)$$

valid for both convex and concave profiles of $\varepsilon(z)$ with $N^2 > 0$ (hence the "+" index for function $\theta_+$, and hereafter in $N_+$). With $\eta_0 = \eta(d)$, one has:

$$\begin{aligned}
\theta_+ &= n_0^2 + \frac{1+\gamma^2/4}{N_+^2} + \left[1 + \frac{\gamma^2}{4} + n_0^2 N_+^2\right]\left[1 + \frac{1}{(qL_2)^2}\left(y - \frac{z}{L_2}\right)^2\right] + \ldots \\
&\quad + \cos[2q(\eta_0 - \eta)]\left\{n_0^2 - \frac{1+\gamma^2/4}{N_+^2} + \left[1 + \frac{\gamma^2}{4} - n_0^2 N_+^2\right]\left[1 - \frac{1}{(qL_2)^2}\left(y - \frac{z}{L_2}\right)^2\right] + \frac{2\gamma n_0 N_+}{qL_2}\left(y - \frac{z}{L_2}\right)\right\} + \ldots \\
&\quad + s_1 \sin[2q(\eta_0 - \eta)]\left\{-\frac{\gamma n_0}{N_+} - \frac{2}{qL_2}\left[1 + \frac{\gamma^2}{4} - n_0^2 N_+^2\right]\left(y - \frac{z}{L_2}\right) + \gamma n_0 N_+\left[1 - \frac{1}{(qL_2)^2}\left(y - \frac{z}{L_2}\right)^2\right]\right\}
\end{aligned} \qquad (26)$$

where $\quad \Delta = 1 + n_0 N_+ - \frac{is_1\gamma}{2} \qquad (27)$

### 4.2. Group velocity of evanescent waves in concave barriers.

If the wave frequency is less than the cut – off frequency, the radiation flux will be transmitted through the film in the tunneling regime. Introducing the notations

$$p = \frac{\omega}{c} n_0 N_-; \quad N_- = \sqrt{u^2 - 1}; \quad u = \frac{\Omega}{\omega} > 1 \qquad (28)$$

one finds[11] that this case can be treated as the one above, with the following substitutions

$$q \to ip; \quad N_+ \to iN_-;$$

$$\cos[2q(\eta_0 - \eta)] \to \operatorname{ch}[2p(\eta_0 - \eta)]; \quad \sin[2q(\eta_0 - \eta)] \to i\operatorname{sh}[2p(\eta_0 - \eta)] \qquad (29)$$

The group velocity of the evanescent wave can be found by substituting in (25) instead of $\theta_+$ the following function

$$\theta_- = n_0^2 - \frac{1+\gamma^2/4}{N_-^2} + \left[1 + \frac{\gamma^2}{4} - n_0^2 N_-^2\right]\left[1 - \frac{1}{(pL_2)^2}\left(y - \frac{z}{L_2}\right)^2\right] + ...$$

$$ch\left[2p(\eta_0 - \eta)\right]\left\{n_0^2 + \frac{1+\gamma^2/4}{N_-^2} + \left[1 + \frac{\gamma^2}{4} + n_0^2 N_-^2\right]\left[1 + \frac{1}{(pL_2)^2}\left(y - \frac{z}{L_2}\right)^2\right] + \frac{2\gamma n_0 N_-}{pL_2}\left(y - \frac{z}{L_2}\right)\right\} - ... \quad (30)$$

$$sh\left[2p(\eta_0 - \eta)\right]\left\{\frac{\gamma n_0}{N_-} + \frac{2}{pL_2}\left[1 + \frac{\gamma^2}{4} + n_0^2 N_-^2\right]\left(y - \frac{z}{L_2}\right) + \gamma n_0 N_-\left[1 + \frac{1}{(pL_2)^2}\left(y - \frac{z}{L_2}\right)^2\right]\right\}$$

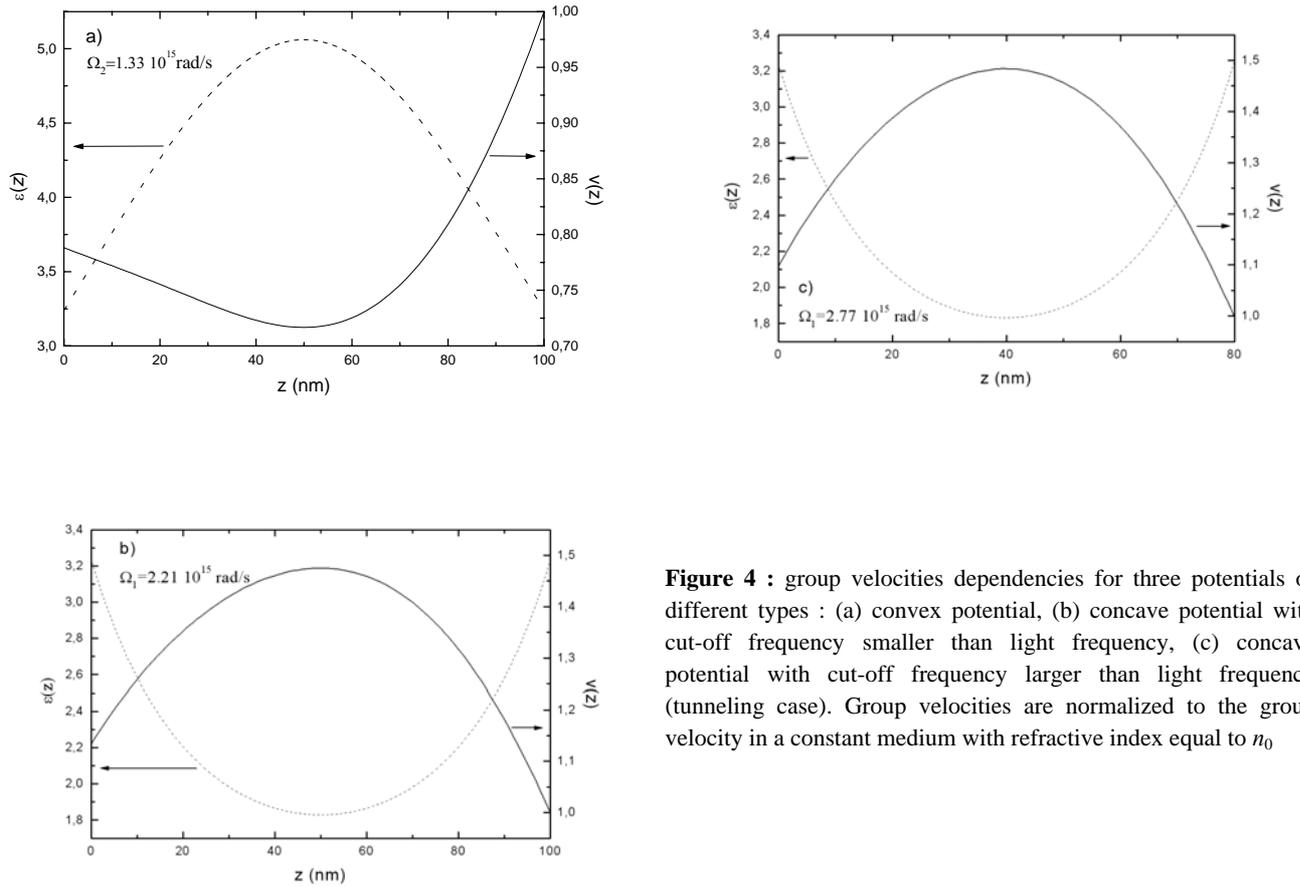

**Figure 4 :** group velocities dependencies for three potentials of different types : (a) convex potential, (b) concave potential with cut-off frequency smaller than light frequency, (c) concave potential with cut-off frequency larger than light frequency (tunneling case). Group velocities are normalized to the group velocity in a constant medium with refractive index equal to $n_0$

The group velocity dependencies corresponding to three types of profiles are represented on figure 4. They are calculated for a wavelength of 800 nm, corresponding to a frequency of $2.36\ 10^{15}$ rad/s. Fig. 4a corresponds to a convex profile, and figs 4 b and c to concave profiles with parameter such that fig 4b corresponds to the case of a propagative transmission, while fig 4c corresponds to the case of a tunneling transmission.

Several remarks can be made concerning the group velocity dependencies displayed on fig 4 :

- despite the fact that the dielectric constant profiles are symmetric, the group velocities are not. This shows that these dependencies are more complex that the simple one expected from the variations of the local index of refraction. analyzing the constituents of the group velocity dependency, one finds that the flux of the Poynting vector is constant while the e.m. power density is not.

- despite the fact that the transmission modes are different between fig 4b (propagating mode) and fig 4c (tunneling modes) the two corresponding dependencies look very similar, while there is a strong difference between the two cases (4a and 4b) corresponding to the propagating mode. Hence it follows that, notwithstanding the propagation mode, it is essentially the characteristics of the dielectric constant profile which dominate the behaviour of the group velocity.
- with the knowledge of the group velocities throughout the three films, it is straightforward to calculate, using eq. (25), and integrating over $z$ between $z=0$ and $z=d$, the group delay induced by the three films. One finds that in all cases this group delay remains subluminal (meaning that it is smaller than $d/c$). Hence, based on this calculation of the group velocity, we find no evidence of superluminal propagation in the case of the tunneling regime. Particularly in the case of tunneling, no phase velocity can be attributed to the evanescent wave in the film (in the case of propagating wave, one could possibly define such a phase velocity in the $\eta$-space, where the waves are sinusoidal, but we did not explore this possibility). However, it is possible to define a phase delay time by calculating the phase shift $\varphi_-$ introduced by the tunneling through the film in the transmission function. This "phase-time" has been abundantly discussed in the literature, and served as a base for the claim of "superluminal propagation", and is defined as

$$t_p = \frac{\partial \varphi_-}{\partial \omega} \qquad (31)$$

Applying this definition, one finds[11] that the phase time corresponding to fig 4c is 0.4 fs, corresponding to an average velocity of 0.65 $c$. No superluminal propagation can be suspected on this ground either. Moreover, it is shown in [11] that there is a link between the phase shifts in transmission and in reflection. The following relation can be established :

$$\varphi_t - \varphi_R = \frac{\pi}{2} \qquad (t_p)_t = -(t_p)_R \qquad (32)$$

This relation offers a possibility of investigating the phase time dependence without being handicapped by the very strong attenuation of the wave in the case of tunneling, and such a study was performed in [20] in the radiofrequency case. This is particularly interesting for studying the Hartman effect, namely the saturation of phase shift $\varphi_-$ when the thickness of the sample increases. In our case, the only way of increasing significantly the film thickness without changing its fundamental property (a cut-off frequency larger than the light frequency) is to stack a number of identical plates. For such a system, a recursive relation can be established for the parameter $Q$ (eq. 24) which writes

$$Q_{m+1} = \exp(-2mp\eta_0)Q_0 \qquad (33)$$

allowing to calculate the phase shift introduced by a set of $m$ films :

$$\varphi_{m-} = \text{Arctg}\left[\frac{\text{th}(mp\eta_0)\left(1 - \gamma^2/4 - n_0^2 N_-^2\right) + \gamma n_0 N_-}{2 n_0 N_- - \gamma \,\text{th}(mp\eta_0)}\right] \qquad (34)$$

as well as the corresponding phase time. Their dependence on the number of films (with the same parameters as on fig 4c) is represented on figure 5.

One sees that both the phase shift and the phase time increases for the first few films (typically $m<6$), but that there is practically no further evolution, and thus a saturation as observed in the Hartman effect. The saturated phase time can be expressed as :

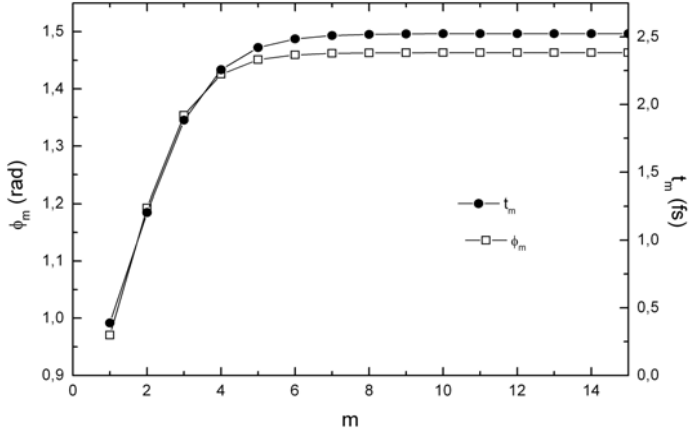

**Figure 5** : illustration of the Hartman effect in a stack of *m* films identical to that of fig 4c : saturation of the transmission dephasing and of the phase time as a function of the number of films in the stack

$$t_p = \frac{2}{\omega}\left[1+\left(\frac{\gamma}{2}-n_0 N_-\right)^2\right]^{-2}\left\{\left[\frac{\gamma^2}{2}+2n_0^2 u^2 - \gamma n_0 N_-\left(1+\frac{u^2}{N_-^2}\right)\right]\left(\frac{\gamma}{2}-n_0 N_-\right)+\frac{n_0 u^2}{N_-}-\frac{\gamma}{2}\right\} \quad (35)$$

so that if the normalized frequency is such that the following condition :

$$\frac{\gamma}{2}-n_0 N_- = 1 \quad (36)$$

is satisfied, $t_p$ has a very simple expression :

$$t_p = \frac{1}{2\omega}\left(1-\frac{n_0}{N_-}\right) \quad (37)$$

which can be negative if $n_0$ is high enough. One can find solutions in $\omega$ such that both conditions (35) and (36) are realized and one is in the tunneling regime. One has to have $\gamma^2(n_0^2-1)>1$, a condition easy to satisfy. This leads us to the conclusion that the phase time, though having the dimension of a time, possesses properties that preclude its acceptation as a physical time, and therefore its use as a ground for the evidence of superluminal propagation. Finally, it can be shown[11] that the saturation of phase time observed on fig. 5 is specific of a lossless medium. Introduction of even a very small absorption in the film (by using a complex index of refraction) leads to a constantly growing phase in the case of figure 5. Moreover, it is possible to generalize as above the calculation of the group velocity to the case of a stack of *m* layers : one finds that for allvalues of *m*,the group velocity stays subluminal.

Finally, we have studied here the intrinsic properties of inhomogeneous films, while as in the case of section 3, such films will be used in the optical domain as coatings. In the case of a semi-infinite substrate with refractive index $n_1$, the modifications to the above equation will result from the change in eq (24), $Q$ becoming :

$$Q = \frac{-\exp(2iq\eta_0)\left(n_1 - \frac{i}{2}s_1\gamma - n_0 N\right)}{n_1 - \frac{i}{2}s_1\gamma + n_0 N} \quad (38)$$

This will affect the amplitude of the wave reflected at the interface $z = d$, and thus the structure of both the Poynting vector and the energy density (the incoming wave in not modified) and, consequently, the group velocity (through the functions $\theta$). The above calculations can easily be reproduced in such a case, and their results do not change the general conclusions given above.

## 5. GROUP VELOCITY DISPERSION IN INHOMOGENEOUS OPTICAL BARRIERS

A last interesting property that can be obtained from the group velocity is the group velocity dispersion introduced by such films. Indeed, we have seen that the HID is much stronger than typical intrinsic material dispersion, and that it is also frequency dependent (in a plasma or waveguide way). One can thus expect that such films will exhibit a strong group velocity dispersion. Here again, one has to investigate this point without resorting to the usual definition with the help of the phase velocity, which is not defined in such films. To do so one can use the group delay time calculated from the group velocity inside the film, as mentioned above. Such a group delay time can be obtained in all cases, but the case of transmitting barriers is the most relevant in terms of possible applications. In such a case, one has to use the "+" definition of the group velocity (eq 25). Noting

$$A_0 = arctg\left[\frac{2y\sqrt{1-y^2}}{1-2y^2}\right] \quad \text{(convex)} \quad \text{or} \quad A_0 = \ln\left(\frac{\sqrt{1+y^2}+y}{\sqrt{1+y^2}-y}\right) \quad \text{(concave)} \tag{39}$$

one obtains the following expression for the group delay time (which is expressed as a function of the "normal" group delay time $t_0 = d/c$) for the convex pofile of $U(z)$:

$$\frac{T}{t_0} = \frac{1}{8y\sqrt{1-y^2}} \left\{ \begin{array}{l} 2A_0\left[n_0^2 + \left(1+\frac{\gamma^2}{4}\right)\big/N_+^2\right] + \frac{2yu^2}{N_+^2\sqrt{1-y^2}}\left(1+\frac{\gamma^2}{4}+n_0^2 N_+^2\right)+... \\ \frac{u}{2N_+}\sin(2q\eta_0)\left[n_0^2 - \left(1+\frac{\gamma^2}{4}\right)\big/N_+^2 + \left(1+\frac{\gamma^2}{4}-n_0^2 N_+^2\right)\left(1+\frac{u^2}{N_+^2}\right)\right]-..... \\ \gamma n_0 u \frac{u^2}{N_+^2}\left[1-\cos(2q\eta_0)\right] - \frac{y}{\sqrt{1-y^2}}\frac{u^2}{N_+^2}\left[\left(1+\frac{\gamma^2}{4}-n_0^2 N_+^2\right)(1+\cos(2q\eta_0))-\gamma n_0 N_+ \sin(2q\eta_0)\right] \end{array} \right\} \tag{40}$$

and in the case of the concave profile :

$$\frac{T}{t_0} = \frac{1}{8y\sqrt{1+y^2}} \left\{ \begin{array}{l} 2A_0\left[n_0^2 + \left(1+\frac{\gamma^2}{4}\right)\big/N_+^2\right] - \frac{2yu^2}{N_+^2\sqrt{1+y^2}}\left(1+\frac{\gamma^2}{4}+n_0^2 N_+^2\right)+... \\ \frac{u}{2N_+}\sin(2q\eta_0)\left[n_0^2 - \left(1+\frac{\gamma^2}{4}\right)\big/N_+^2 + \left(1+\frac{\gamma^2}{4}-n_0^2 N_+^2\right)\left(1-\frac{u^2}{N_+^2}\right)\right]-..... \\ \gamma n_0 u \frac{u^2}{N_+^2}\left[1-\cos(2q\eta_0)\right] + \frac{y}{\sqrt{1+y^2}}\frac{u^2}{N_+^2}\left[\left(1+\frac{\gamma^2}{4}-n_0^2 N_+^2\right)(1+\cos(2q\eta_0))+\gamma n_0 N_+ \sin(2q\eta_0)\right] \end{array} \right\} \tag{41}$$

Figure 6 presents the particularly interesting case of a concave profile of $U$ used just above its cut-off frequency. In expression (41), many quantities are extremely frequency dependent close to the cut-off frequency and so does the group delay time, as shown on figure 6. In the usual language of group velocity dispersion, this case corresponds to a negative dispersion (the blue frequencies go faster than the red ones). It should be pointed out that important group delays (as high as 30 fs) are obtained in this case with a film of thickness equal to 100 nm only, which illustrates the very unique properties of HID.

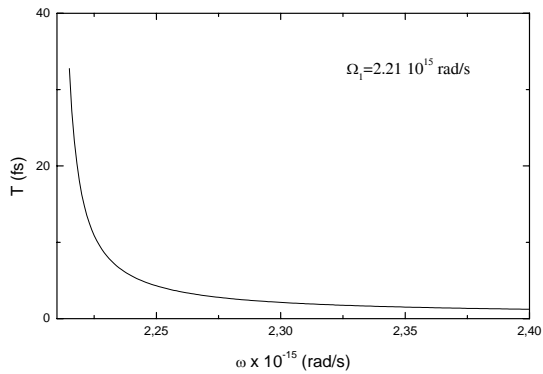 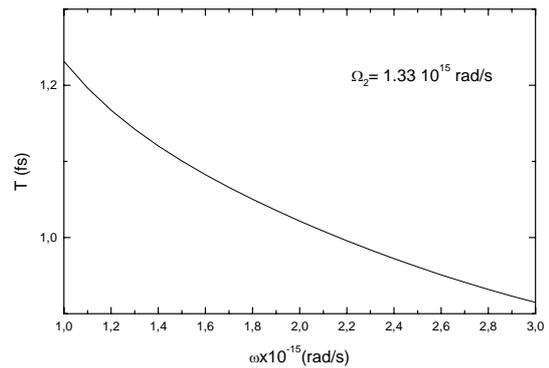

**Figure 6** : group delay time as a function of frequency for a concave index profile (identical to that of figure 4b) used just above its cut-off frequency

**Figure 7** : group delay time as a function of frequency, plotted over a large frequency interval encompassing the film's characteristic frequency in the case of the convex profile of fig 4a

It is enlightening to compare this case to that of a convex profile of $U$ such as the one investigated above (fig 4a). The result of an equivalent calculation over a broad range of frequency encompassing the characteristic frequency shows that contrary to the above case, the group delay stays very small (~ 1fs) with a moderate dispersion, showing no resonant character. This obviously results from the fact that the crossing of the characteristic frequency does not give rise to any resonant phenomena (like, in the concave case, a value of zero for $N_+$ which appears several times in the denominator).

## 6. CONCLUSIONS

In this work, we presented a number of properties which can be studied with the help of exacts solutions of electromagnetic propagation equations, applied to media presenting fast and deep spatial variations of their dielectric constant. Even though this method applies easily to specific analytical forms of such profiles, they contain enough free parameters to describe films with very different structures. Being exact, such solutions are subject to no limitations concerning the amplitude or the length scales over which the dielectric function varies. Several previously investigated cases can be considered as limiting cases for our calculation. Being scalable, this method applies to any wavelength range, from the RF to the XUV, provided that the materials allowing to realize the dielectric constant profiling exist.

Our method allows to understand in detail how the shape of the dielectric constant variation influences the various properties of such films. It reveals the essential concept of Heterogeneity Induced Dispersion (HID), which was shown to be dependent only upon the two spatial scales defining the studied profiles. In the absence of any standard material dispersion, it can lead to effects exceeding by orders of magnitude those generally observed on dispersive materials, and this in the absence of any absorption (which is usually associated with strong dispersive properties, which are generally obtained only close to material's resonances). Depending on the type of profile used (concave or convex), HID can present or not a cut-off frequency below which propagation through the medium is impossible, and evanescent waves are formed.

We studied the reflection properties of such films and observed that they can possess dichroic or antireflection properties, and this over a wide wavelength range. This offers for instance the opportunity of realizing antireflection coatings in domains where standard methods are difficult to apply.

This method also allows to study in detail the propagation or the tunneling of waves through such index-gradient films. In the case of tunneling, we could obtain the associated group delay time and found no sign of superluminal propagation as far as group velocity is concerned. The "phase time" corresponding to such films can also be calculated, and again showed no sign of superluminality, but instead some features – such as the fact that it can take negative values – which forbid to consider it as an acceptable physical time, despite its dimension. However, stacking a variable number of identical films, we were able to reproduce the basic feature giving rise to the Hartman effect, i.e. the saturation of the phase shift when the thickness of the tunneling barrier increases. Finally we showed that when used close to their cut-off frequency, in the propagative mode, such films can have some exceptional negative group velocity dispersion. Trying to otpimize their properties with the aim of manipulating the shape of broadband ultrashort pulses is certainly one of the perspectives opened by this study.

Concerning the possible applications of the results presented above, index-gradient films are currently in use in the optical domain. However, our analytical method allows simple and fast calculations which could prove useful in the monitoring of the growth of such films. One could for instance study algorithms aiming at correcting on line the parameters of a reactor to compensate the deviations measured during the growth. Also the possibility of designing system with specific dispersion or reflection properties in any wavelength range appears as an exciting possibility.

## ACKNOWLEDGEMENTS

The authors acknowledge the support of NATO grant n° PST.CLG.980334